\documentclass{eptcs}
\providecommand{\event}{TFPIE 2024}

\usepackage{amsmath,amssymb}
\usepackage{fontaxes}
\usepackage{graphicx}
\usepackage{stfloats}
\usepackage{multido}
\usepackage{setspace}
\usepackage{textcomp}
\usepackage{alltt}
\usepackage{textcomp}
\usepackage{pgfplots}
\pgfplotsset{compat=newest}
\usepackage{pgfplotstable}
\usepackage{cleveref}
\usepackage{doi}
\usepackage{float}
\floatstyle{ruled}
\restylefloat{figure}
\usepackage{caption}
\usepackage{subcaption}

\newcommand{\racket}{\texttt{Racket}}
\newcommand{\arrow}{\(\rightarrow\)}

\newcommand{\fsm}{\texttt{FSM}}
\newcommand{\sig}{\texttt{\(\Sigma\)}}

\newcommand{\dotss}{\(\ldots\)}

\newcommand{\quot}{\texttt{\textquotesingle{}}}
\newcommand{\dquot}{\texttt{"}}
\newcommand{\qquot}{\texttt{\textasciigrave{}}}
\newcommand{\ndfa}{\texttt{ndfa}}
\newcommand{\dfa}{\texttt{dfa}}
\newcommand{\delt}{\texttt{\(\delta\)}}
\newcommand{\flatt}{\texttt{FLAT}}
\newcommand{\regexp}{\texttt{regexp}}
\newcommand{\jflap}{\texttt{JFLAP}}
\newcommand{\oflap}{\texttt{OpenFlap}}
\newcommand{\gnfa}{\texttt{GNFA}}
\newcommand{\gviz}{\texttt{Graphviz}}

\title{Finite-State Automaton To/From Regular Expression Visualization}

\author{Marco T. Moraz\'an and Tijana Minić
\institute{Seton Hall University}
\email{\{morazanm|minictij\}@shu.edu}}

\def\titlerunning{Finite-State Automaton To/From Regular Expression Visualization}
\def\authorrunning{Moraz\'an and Minić}

\begin{document}

\maketitle

\begin{abstract}
Most Formal Languages and Automata Theory courses explore the duality between computation models to recognize words in a language and computation models to generate words in a language. For students unaccustomed to formal statements, these transformations are rarely intuitive. To assist students with such transformations, visualization tools can play a pivotal role. This article presents visualization tools developed for \fsm{}--a domain-specific language for the Automata Theory classroom--to transform a finite state automaton to a regular expression and vice versa. Using these tools, the user may provide an arbitrary finite-state machine or an arbitrary regular expression and step forward and step backwards through a transformation. At each step, the visualization describes the step taken. The tools are outlined, their implementation is described, and they are compared with related work. In addition, empirical data collected from a control group is presented. The empirical data suggests that the tools are well-received, effective, and learning how to use them has a low extraneous cognitive load.
\end{abstract}

\section{Introduction}

Formal Languages and Automata Theory (\flatt{}) courses emphasize the equivalence of different computation models. For instance, the equivalence of deterministic finite-state machines (\dfa{}s) and nondeterministic finite-state machines (\ndfa{}s) is established by showing students how to transform an \ndfa{} into a \dfa{}. Such a transformation, although not trivial for a first-time \flatt{} student, is relatively intuitive: a machine (i.e., an \ndfa{}) to recognize words in a language is transformed into a different machine (i.e., a \dfa{}) to recognize words in the same language. In essence, an algorithm to determine language membership is transformed into another algorithm to determine language membership. 

Less intuitive transformations are those from a model that recognizes words in a language to a model that generates words in a language and vice versa. That is, an algorithm to recognize words in a language is transformed into an algorithm to generate words in the same language and an algorithm to generate words in a language is transformed into an algorithm to recognize words in the same language. In such cases, the generated algorithm does not satisfy the same purpose. In a typical \flatt{} course, for instance, students learn how to transform a pushdown automaton into a context-free grammar and vice versa and learn how to transform a regular grammar into a finite-state automaton and vice versa.

Transformations from a generating algorithm to a recognizing algorithm and from a recognizing algorithm to a generating algorithm may be confusing for first-time \flatt{} students given that formal statements are rarely intuitive for them. Chief among these transformations is the one from an \ndfa{} to a regular expression (\regexp) and back. These transformations are important, because \regexp{}s are well-suited for humans to express patterns and finite-state automata are well-suited for program development \cite{Gruber}. For example, \regexp{}s are used in tools such as \texttt{awk} \cite{Robbins} and \texttt{emacs} \cite{Emacs} while finite-state automata are at the heart of algorithms for string searching \cite{KMP} and lexical analysis \cite{Perrin}. To aid student understanding, visualization tools like \jflap{} \cite{Rodger,RodgerII} and \oflap{} \cite{Mohammed} have been developed. It is usually assumed that visualizations are a powerful pedagogic tool in the classroom. They allow students to interact, to some degree or another, with their designs. This, however, may not be enough to have an effective teaching tool for several reasons. A visualization requires users to learn its interface and this can place an extraneous cognitive load on students \cite{Hegarty,Sweller}. To be effective and reduce such a load, visualizations must provide representations that behave as the objects themselves \cite{Hutchins}. This does not mean that a visualization tool cannot offer more advanced features. It means that it is important for there to be some easy-to-use features.

This article describes the visualization tools developed for \fsm{} \cite{fsm} to aid student understanding of the transformations from an \ndfa{} to a \regexp{} and from a \regexp{} to an \ndfa{}. \fsm{} is a functional domain-specific language embedded in \racket{} \cite{Racket} developed for the \flatt{} classroom to program state machines, grammars, and regular expressions \cite{PBFLAT}. The tools have multiple goals that include aiding in the understanding of the construction algorithms, reducing the extraneous cognitive load, and allowing students to examine construction steps interactively both forward and backwards. The article is organized as follows. \Cref{rw} reviews and contrasts related work. \Cref{fsm-intro} presents a brief introduction to the \fsm{} syntax needed to navigate this article. \Cref{viz-design} discusses the overall design idea behind the visualization strategies. \Cref{impl} outlines the generation of visualization graphics. \Cref{study} presents empirical data collected, in preparation for classroom deployment, from a control group. Finally, \Cref{concls} delivers concluding remarks and discusses directions for future work.

\section{Related Work}
\label{rw}

\subsection{Construction Algorithms}

\subsubsection{\regexp{} to \ndfa{}}

Let $\Sigma$ be the alphabet of a language. There are six regular expression varieties \cite{Sipser}:
\begin{enumerate}
\item R = \texttt{a}, where a$\in \Sigma$

\item R = $\epsilon$, where $\epsilon$ denotes the empty word

\item R = $\varnothing$, denotes the empty language

\item R = R$_1 \cup$ R$_2$, denotes the union of two regular expressions

\item R = R$_1 \circ$ R$_2$, denotes the concatenation of two regular expressions

\item R = R$_1^*$, denotes zero or more concatenations of a regular expression

\end{enumerate}
The creation of an \ndfa{} for varieties 1--3 is straightforward. For varieties 1--2, each corresponding \ndfa{} has a starting state and a final state. The transition between these consumes an alphabet element for the first variety and nothing (i.e., the empty string) for the second variety. The \ndfa{} for the third variety only has a starting state and no final state. The transformation for varieties 4--6 hinges on closure properties for regular languages. That is, an \ndfa{} is constructed using the algorithms developed as part of the constructive proofs establishing that the languages accepted by \ndfa{}s are closed under union, concatenation, and Kleene star. For union and concatenation, \ndfa{}s M$_1$ and M$_2$ are recursively constructed for R$_1$ and R$_2$. For union, a new starting and a new final state are created. The resulting \ndfa{} nondeterministically moves from the new starting state to either  M$_1$'s or M$_2$'s starting state and from each of M$_1$'s and M$_2$'s final states to the new final state. For concatenation, nondeterministic transitions are added from M$_1$'s final states to M$_2$'s starting state and the new machine's final states are M$_2$'s final states. For Kleene star, an \ndfa{}, M$_1$, is recursively constructed for \texttt{R$_1$}. A new starting state, that is also a final state, is generated. In addition, nondeterministic transitions from the new start state to M$_1$'s start state and from M$_1$'s final states to M$_1$'s start state are generated. The reader may consult any introductory \flatt{} textbook for the formal details of these constructors (e.g., \cite{Hopcroft,Lewis,Linz,Martin,PBFLAT,Rich,Sipser}.

A transformation is commonly explained using a generalized nondeterministic finite-state automata (\texttt{GNFA}) \cite{Sipser}. A \texttt{GNFA} is similar to an \ndfa{}, but its transitions are done on regular expressions. The initial \texttt{GNFA} has two states and a transition between them. The transition is on the regular expression that is being transformed. We have chosen this approach for our visualization tools, because at each step a single compound regular expression (i.e., union, concatenation, or Kleene star) may be decomposed to create the necessary new sub-\texttt{GNFA}s. The focus on a single edge facilitates the generation of an informative message that explains the step taken and, thus, reduces the extraneous cognitive load for students.

\subsubsection{\ndfa{} to \regexp{}}

\flatt{} textbooks usually outline the \ndfa{} to \regexp{} transformation using either an elegant set of recursive equations or a graph-based approach using the \ndfa{}'s transition diagram. The equation-based approach represents the language of the machine constructed as the union of a finite number of small languages \cite{Lewis}. By numbering the machine states \texttt{K}=\{k$_{\texttt{1}}$, k$_{\texttt{2}}$, \dotss{}, k$_{\texttt{n}}$\}, where \texttt{k$_{\texttt{1}}$} is the starting state, the regular expression for all words that take the machine from state \texttt{k$_{\texttt{i}}$} to state \texttt{k$_{\texttt{j}}$} without traversing a state numbered \texttt{m+1} or greater is denoted by \texttt{R(i,j,m)}\footnote{An intermediate state is denoted as \texttt{k$_{\texttt{r}}$}, such that \texttt{0$\leq$r$\leq$m}. The number of intermediate states is not relevant.}. Therefore, we have that the regular expression for the language of an \ndfa{}, \texttt{N}, with \texttt{n} states is constructed as follows:
\begin{alltt}
     L(N) = \(\bigcup\)\{R(1,j,n) | k\(\sb{\texttt{j}}\in{}\)F\}, where F is N's set of final states
\end{alltt}
That is, \texttt{N}'s language contains all words that take the machine from the starting state to a final state by traversing any state. Assuming $\Delta$ is the given machine's set of transition rules, the regular expression is constructed using the following algorithm:
\[ R(i,j,n)=
   \begin{cases}
      \{a|(k_i \ a \ k_j)\in\Delta\} & \text{if } n=0 \wedge i\neq{}j \\
      \{\epsilon\} \bigcup \{a|(k_i \ a \ k_j)\in\Delta\} & \text{if } n=0 \wedge i=j \\
      R(i,j,n-1) \bigcup R(i,n,n-1)R(n,n,n-1)^*R(n,j,n-1) & \text{if } n\neq0 \\
   \end{cases}
\]
This recursive equation states that two cases are distinguished when there can be no intermediate states traversed (i.e., n=0). The first is when $k_i\neq{}k_j$. In this case, we have that all the singletons consumed by rules that directly transition from $k_i$ to $k_j$ are the needed regular expressions. The second is when $k_i=k_j$. In this case, $\epsilon$ is added to the set of symbols consumed on a self loop. If intermediate states may be traversed (i.e., n$\neq$0) then the needed regular expression is the union of two regular expressions. The first generates all words that take the machine from $k_i$ to $k_j$ without traversing a state numbered \texttt{n} or greater. The second concatenates three regular expressions: one that generates all words that take the machine from \texttt{$k_i$} to \texttt{$k_n$} without traversing a state greater than \texttt{n-1}, one that generates all words that take the machine from \texttt{$k_n$} to \texttt{$k_n$} an arbitrary number of times without traversing a state greater than \texttt{n-1}, and one that generates all words that take the machine from \texttt{$k_n$} to \texttt{$k_j$} without traversing a state greater than \texttt{n-1}.

The graph-based approach converts the \ndfa{}, \texttt{N}, into a \texttt{GNFA} and then converts the \texttt{GNFA} to a regular expression \cite{Sipser}. The transformation may be visualized as performing surgery on a directed graph. A \texttt{GNFA} is built starting with \texttt{N}'s transition diagram and adding a new start state, a new final state, and empty transitions from the new start state to \texttt{N}'s start state and from \texttt{N}'s final states to the new final state. In addition, there is a single edge in each direction between any pair of states $k_i$ and $k_j$. If there are one or more transitions from $k_i$ to $k_j$ then the label on the edge in the \texttt{GNFA} is a union-\regexp{} containing a singleton-\regexp{} for each of these transitions. Finally, if there are no edges from $k_i$ to $k_j$ then the label of the arrow is a null-\regexp{}. Such added transitions do not change \texttt{N}'s language because they represent hypothetical transitions and can never be used. The regular expression's computation proceeds by ripping out nodes piecemeal until only the new start state and the new final state remain. At this point, the regular expression on the only remaining edge is for the language of \texttt{N}. When a node $k_r$ is ripped out, the edges into $k_r$, \texttt{($k_i$ a $k_r$)}, and the edges out of $k_r$, \texttt{($k_r$ b $k_j$)} are replaced. If $k_r$ does not have a self-loop then \texttt{($k_i$ a $k_r$)} and \texttt{($k_r$ b $k_j$)} are replaced with \texttt{($k_i$ ab $k_j$)}. If there is a self-loop on $k_r$ then \texttt{($k_r$ b $k_j$)}, \texttt{($k_r$ c $k_r$)}, and \texttt{($k_r$ b $k_j$)} are replaced by \texttt{($k_i$ ac$^*$b $k_j$)}. Finally, after ripping out $k_r$, multiple edges between nodes are replaced with a single edge that is labeled with the union of labels of the multiple edges.

\fsm{}'s visualization uses a graph-based approach, because it tends to be easier to understand by first-time \flatt{} students. The small local step approach of ripping out one node at a time is more palatable than, for example, computing \texttt{R(i,j,n)} which requires a more global view of the transition relation. The algorithm used, like the algorithm described by Sipser \cite{Sipser}, builds a \texttt{GNFA} by adding new a new start state, a new final state, and the corresponding empty transitions. In contrast, however, the addition of edges with a null-\regexp{} label is suppressed. Such edges serve no real purpose and, for visualization purposes, clutter the graphics produced. Instead of adding such edges, when a node is ripped out its predecessors and its successors are computed to properly substitute the edges into and out of the ripped out node. In this manner, the graphs produced are easier to read by students and lower the extraneous cognitive load.

\subsection{Visualization}

\begin{figure}[t!]
\captionsetup[subfigure]{justification=centering}
\begin{subfigure}[b]{0.48\textwidth}
\centering
\includegraphics[scale=0.5]{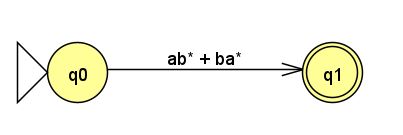}
\caption{\texttt{GNFA} for \texttt{L=\{$ab^{\texttt{*}} \cup ba^{\texttt{*}}$\}}.}
\label{abs-bas-ndfa-graph}
\end{subfigure}
\begin{subfigure}[b]{0.48\textwidth}
\centering
\includegraphics[scale=0.5]{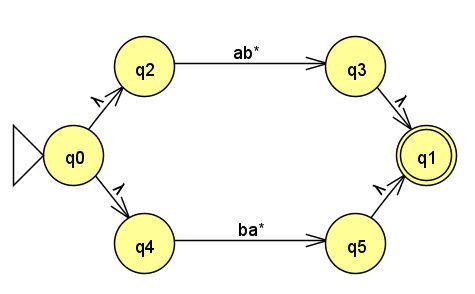}
\caption{\gnfa{} after one step.}
\label{abs-bas-ndfa-graph-s2}
\end{subfigure}
\caption{Initial step in the \jflap{} visualization.} \label{jflap-init-steps}
\end{figure}

\jflap{} is a visualization tool that supports the \ndfa{} to \regexp{} transformation and vice versa. To transform from an \ndfa{} to a \regexp{}, the user must manually construct the \ndfa{}. This includes graphically drawing nodes and edges, marking the starting and final states, and laying out the transition diagram in an appealing manner. To transform from a \regexp{} to an \ndfa{}, the user must use a concrete grammar to write the regular expression. This grammar uses \texttt{+} for union, \texttt{*} for Kleene star, \texttt{!} for the empty word, and parentheses to define the order of operations. For neither conversion is the user given the ability to step back through the computation to review previous steps.

\subsubsection{\regexp{} to \ndfa{} Visualization}

\begin{figure}[t!]
\captionsetup[subfigure]{justification=centering}
\centering
\begin{subfigure}[b]{\textwidth}
\centering
\includegraphics[scale=0.6]{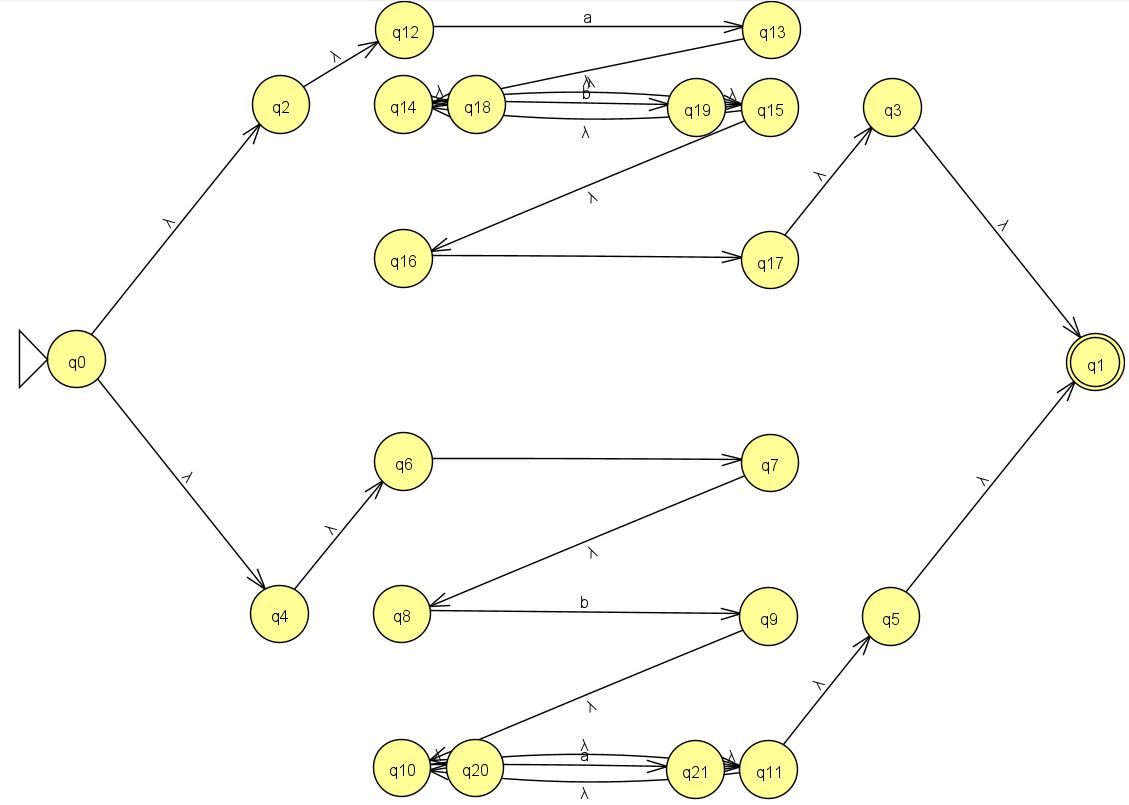}
\caption{\jflap{} transition diagram.}
\label{jflap-final}
\end{subfigure}
\\
\begin{subfigure}[b]{\textwidth}
\centering
\includegraphics[scale=0.45]{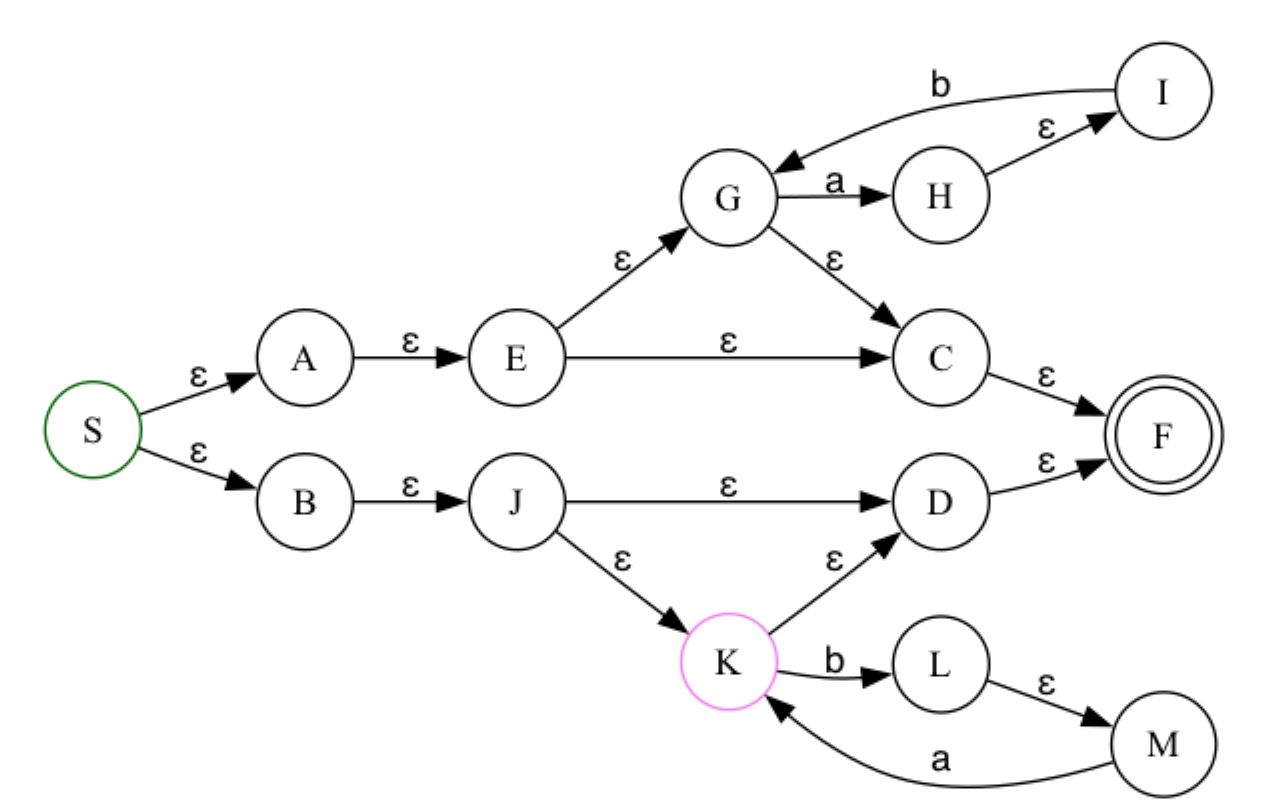}
\caption{\fsm{} transition diagram.}
\label{fsm-final}
\end{subfigure}
\caption{Resulting transition diagrams for \texttt{L=\{$ab^{\texttt{*}} \cup ba^{\texttt{*}}$\}}.} \label{final-ndfas}
\end{figure}

The conversion is done from a \gnfa{} to \regexp{}. At each step, a decomposable regular expression is transformed using \ndfa{} closure properties over union, concatenation, and Kleene star. Such a step may be manually done by the user or may be done automatically by pressing a \texttt{Do Step} button. To illustrate how a step is done, consider the \gnfa{} in \Cref{abs-bas-ndfa-graph} for \texttt{L=\{$ab^{\texttt{*}} \cup ba^{\texttt{*}}$\}}\footnote{The nodes have been manually rearranged to make the illustrations easy to read.}. The result of performing the first transformation step results in the \gnfa{} displayed in \Cref{abs-bas-ndfa-graph-s2}. The edge from \texttt{q0} to \texttt{q1} is substituted with a \gnfa{} that starts at \texttt{q0}, nondeterministically transitions to a (sub-)\gnfa{} for one of the union's branches, and from both (sub-)\gnfa{}s nondeterministically transitions to \texttt{q1}. The conversion process may continue piecemeal or may be completed in one step by pressing a \texttt{Do All} button that results, without manually rearranging nodes to improve readability, in the \ndfa{} displayed in \Cref{jflap-final}. As the reader can appreciate, nodes are haphazardly placed and some edges are impossible to read, thus, requiring the user to rearrange nodes to make the transition diagram readable.

\fsm{}'s visualization also uses a graph-based approach to generate the \ndfa{}. Like \jflap{}, closure properties of regular languages over union, concatenation, and Kleene star are used to transform decomposable \regexp{}s. In contrast, however, a primary goal of the \fsm{} visualization is to reduce the extraneous cognitive load. To this end, the tool always selects the next decomposable \regexp{} to transform and allows the user to step back in the computation to review transformation steps. In addition, an informative message is displayed highlighting the edge that is transformed. In this manner, the user does not have to wonder what occurred when a step is not clear to them. In further contrast with \jflap{}, every transition diagram is rendered using \gviz{} \cite{gviz1,gviz2}. Thus, nodes and edges are rendered in an appealing manner that makes reading the rendered transition diagram easier. For instance, compare the transition diagram layout obtained using \fsm{}'s visualization displayed in \Cref{fsm-final} with its counterpart obtained using \jflap{} in \Cref{jflap-final}.

\subsubsection{\ndfa{} to \regexp{}}

\begin{figure}[t!]
\centering
\includegraphics[scale=0.5]{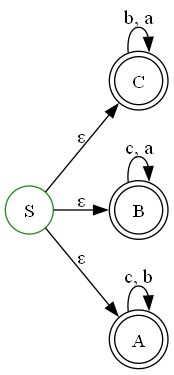}
\caption{An \texttt{ndfa} for the language L=\{w $|$ w is missing at least one in \{a b c\}\}.}
\label{at-least-one-missing}
\end{figure}

\begin{figure}[t!]
\captionsetup[subfigure]{justification=centering}
\centering
\begin{subfigure}[b]{0.4\textwidth}
\centering
\includegraphics[scale=0.35]{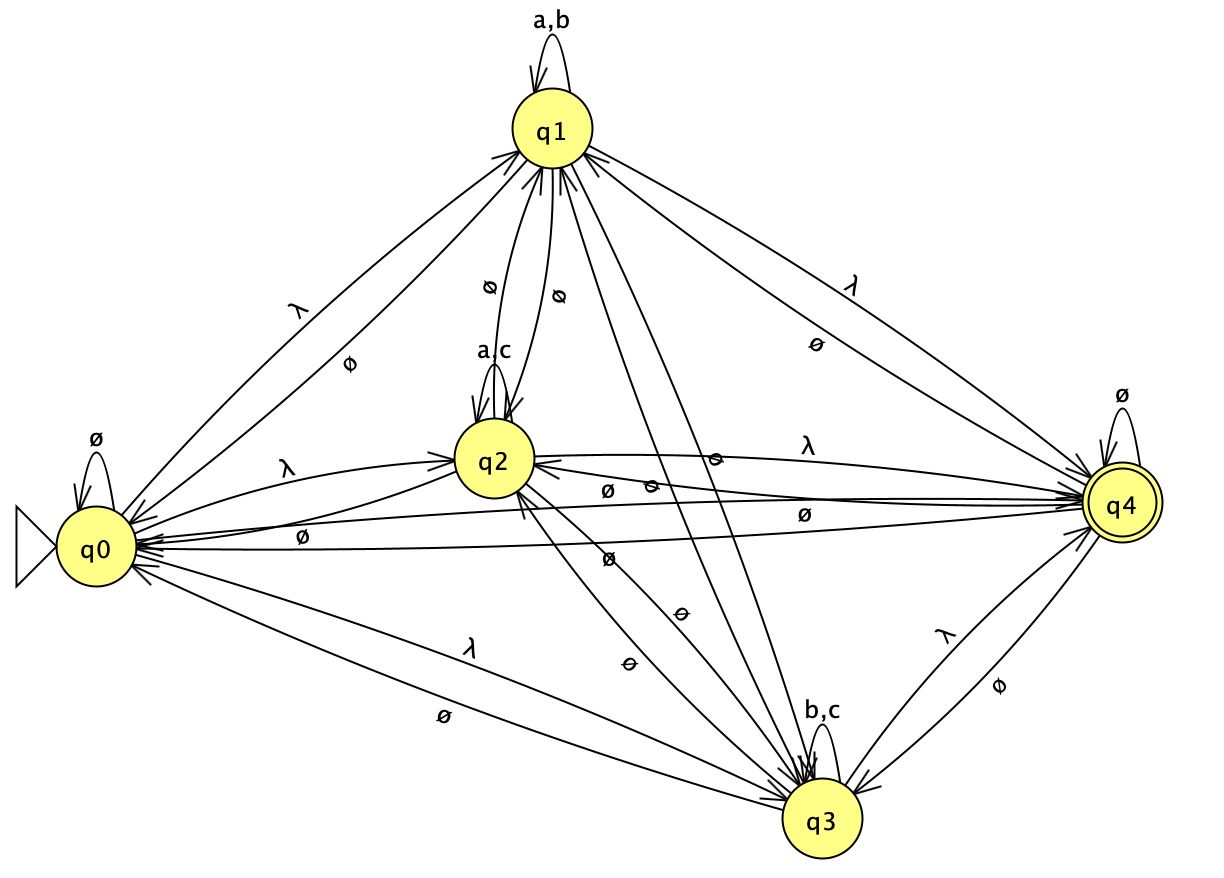}
\caption{Visualization state after adding null-labeled edges.}\label{step3jflap}
\end{subfigure}
\hfill
\begin{subfigure}[b]{0.5\textwidth}
\centering
\includegraphics[scale=0.35]{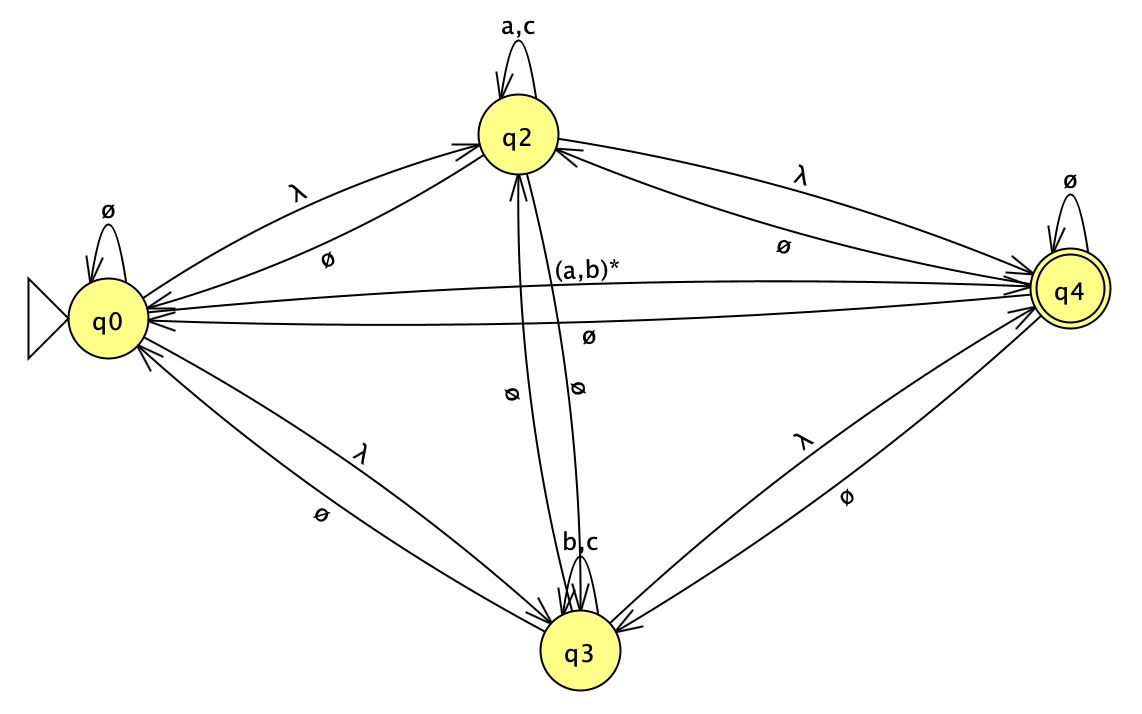}
\caption{Visualization state after ripping out \texttt{q1}.} \label{ripped-node}
\end{subfigure}
\caption{\texttt{JFLAP} visualization states.}
\end{figure}

\begin{figure}[t!]
\captionsetup[subfigure]{justification=centering}
\centering
\begin{subfigure}[b]{0.45\textwidth}
\centering
\includegraphics[scale=0.75]{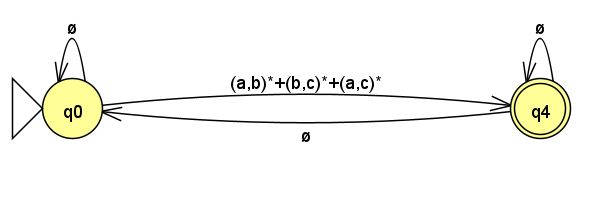}
\caption{\texttt{JFLAP}'s final visualization state.}
\label{jflapcomplete}
\end{subfigure}
\hfill
\begin{subfigure}[b]{0.45\textwidth}
\centering
\includegraphics[scale=0.40]{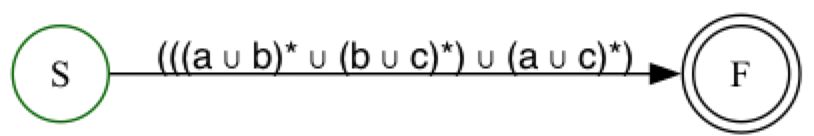}
\caption{\fsm{}'s final visualization state.}
\label{fsmcomplete}
\end{subfigure}
\caption{\texttt{JFLAP}'s and \fsm{}'s final visualization state.}
\end{figure}

\jflap{}'s \ndfa{} to \regexp{} transformation visualization follows the graph-based approach described above. The user manually performs each step of the transformation by following provided instructions. The instructions have the user add a new dead state, create a \gnfa{} by combining edges with multiple labels into a union \regexp{}, add null-labeled directed edges between states that do not have an edge between them, and, finally, rip out nodes. The user has the option to complete each of these steps manually or automatically. To illustrate the conversion, consider transforming the \ndfa{} displayed in \Cref{at-least-one-missing}. \Cref{step3jflap} displays the visualization's state after adding a new final state and the necessary null transitions between states (nodes have been moved to improve readability). Despite moving nodes to improve readability, we can observe that the visualization's state is hard to read at best. The primary problem is that the visualization is cluttered with null-labeled edges that result in edge overlapping. \Cref{ripped-node} displays the visualization state after ripping out the first node (\texttt{q1} in this example). We can once again observe that the visualization state is hard to read. In addition, it is difficult to visually discern the effect of ripping out a node. Finally, \Cref{jflapcomplete} displays \jflap{}'s final visualization state. In the final state, it becomes easy to discern the resulting regular expression despite the (useless) null-labeled transitions.

As in \jflap{}, \fsm{}'s visualization also follows the graph-based transformation approach. In contrast, however, the user is much less burdened. The user does not have to manually add a final state, create a \gnfa{}, add null-labeled directed edges, nor choose the next node to rip out. All this is automatically done or omitted as the user steps through the visualization. Thus, reducing the extraneous cognitive load. The user only needs to step forward and backward between visualization states using the arrow keys. The ability to step backwards in the transformation allows users to visually observe how edges are combined when a node is ripped out. Finally, every transition diagram is rendered using \gviz{} \cite{gviz1,gviz2} to provide an appealing layout. For instance, for the \ndfa{} displayed in \Cref{at-least-one-missing}, the \fsm{} visualization's final state is rendered as displayed in \Cref{fsmcomplete}. The reader can appreciate that the graphic is more appealing than the graphic produced by \jflap{}.

\section{A Brief Introduction to \fsm}
\label{fsm-intro}

\fsm{} is a domain specific language, embedded in \racket{}, for the \flatt{} classroom. In \fsm{}, state machines, grammars, and regular expressions are first-class. Nondeterminism is a built-in language feature that programmers may use as they use their favorite features in any programming language. The \fsm{} types relevant for this article are regular expressions and finite-state machines. 


\subsection{Regular Expressions}

\begin{figure}[t!]
\begin{alltt}
     #lang fsm

     (define a (singleton-regexp "a"))     (define b (singleton-regexp "b"))

     (define a* (kleenestar-regexp a))     (define b* (kleenestar-regexp b))
     
     (define ab* (concat-regexp a b*))     (define ba* (concat-regexp b a*))

     ;; L= ab* U ba*
     (define ab*Uba* (union-regexp (concat-regexp a b*) (concat-regexp b a*)))

     ;; word \arrow{} Boolean
     ;; Purpose: Determine if the given word is in ab* U ba*
     (define (in-ab*Uba*? w)
       (or (and (eq? (first w) \quot{}a) (andmap (\(\lambda\) (s) (eq? s \quot{}b)) (rest w)))
           (and (eq? (first w) \quot{}b) (andmap (\(\lambda\) (s) (eq? s \quot{}a)) (rest w)))))

     (check-pred in-ab*Uba*? (gen-regexp-word ab*Uba*))
     (check-pred in-ab*Uba*? (gen-regexp-word ab*Uba*))
     (check-pred in-ab*Uba*? (gen-regexp-word ab*Uba*))
\end{alltt}
\caption{The \fsm{} regular expression for \texttt{L=\{$\texttt{ab}^{\texttt{*}} \cup \texttt{ba}^{\texttt{*}}$\}}.} \label{regexp-fsm-ex}
\end{figure}

The constructors for a regular expression, over an alphabet \sig, are:
\begin{enumerate}
    \item (null-regexp)
    \item (empty-regexp)
    \item (singleton-regexp \dquot{}a\dquot{})\textrm{, where a\(\in\)\(\Sigma\)}
    \item (union-regexp r1 r2)\textrm{,  where r1 and r2 are regular expressions}
    \item (concat-regexp r1 r2)\textrm{, where r1 and r2 are regular expressions}
    \item (kleenestar-regexp r1)\textrm{, where r is a regular expression}
\end{enumerate}
The \fsm{} selector functions for regular expressions are:
\begin{quote}
\begin{description}
  \item[singleton-regexp-a:] Extracts the embedded string
  \item[kleenestar-regexp-r1:] Extracts the embedded regular expression
  \item[union-regexp-r1:] Extracts the first embedded regular expression
  \item[union-regexp-r2:] Extracts the second embedded regular expression
  \item[concat-regexp-r1:] Extracts the first embedded regular expression
  \item[concat-regexp-r2:] Extracts the second embedded regular expression
\end{description}
\end{quote}
The following predicates are defined to distinguish among the regular expression subtypes:
\begin{alltt}
     empty-regexp?     singleton-regexp?     kleenestar-regexp?
     union-regexp?     concat-regexp?        null-regexp?
\end{alltt}
Each consumes a value of any type and returns a Boolean. Finally, the observer, \texttt{gen-regexp-word}, takes as input a regular expression and returns a word in the language of the given regular expression. This observer nondeterministically decides how many repetitions of a \texttt{kleenestar-regexp} to generate and nondeterministically decides which branch of a \texttt{union-regexp} to use in generation.

As a programming example, consider the \fsm{} regular expression displayed in \Cref{regexp-fsm-ex} for \texttt{L=\{$\texttt{ab}^{\texttt{*}} \cup \texttt{ba}^{\texttt{*}}$\}}. The code is made readable by independently defining each needed sub-\regexp{}. The reader can appreciate that this makes the implementation accessible for almost any student. The unit tests use the auxiliary predicate, \texttt{in-ab*Uba*?}, to determine if a generated word is in \texttt{L}. The tests all look the same, but they are not (in all likelihood) given that each word generation, as described above, nondeterministically decides the number of repetitions for a Kleene star and the branch of the union used to generate a word.

\subsection{Finite-State Automatons}

The \fsm{} machine constructors of interest for this article are those for finite-state automata:
\begin{alltt}
     make-dfa:  K \sig{} s F \delt{} \arrow{} dfa     make-ndfa: K \sig{} s F \delt{} \arrow{} ndfa
\end{alltt}
\texttt{K} is a list of states, \sig{} is a list of alphabet symbols, \texttt{s}$\in$\texttt{K} is the starting state, \texttt{F}$\subseteq$\texttt{K} is a list of final states, and \texttt{$\delta$} is a transition relation (that must be a function for a \dfa{}). A transition relation is represented as a list of transition rules. A \dfa{} transition rule is a triple, \texttt{(K $\Sigma$ K)}, containing a source state, the element to read, and a destination state. For an \ndfa{} transition, the element to read may be \texttt{EMP} (i.e., nothing is read).

The observers are:
\begin{alltt}
     (sm-states M) (sm-sigma M)   (sm-start M) (sm-finals M) (sm-rules M)
     (sm-type M)   (sm-apply M w) (sm-showtransitions M w)
\end{alltt}
The first 5 observers extract a component from the given state machine, \texttt{sm-type} returns the given state machine's type, \texttt{sm-apply} applies the given machine to the given word and returns \texttt{\quot{}accept} or \texttt{\quot{}reject}, and \texttt{sm-showtransitions} returns a trace of the configurations traversed when applying the given machine to the given word ending with the result. A trace is only returned, however, if the machine is a \dfa{} or if the word is accepted by an \ndfa{}.

Finally, \fsm{} provides machine rendering and machine execution visualization. The  visualization primitives are:
\begin{alltt}
     (sm-graph M)     (sm-visualize M [(s p)\(\sp{*}\)])
\end{alltt}
The first returns an image for the given machine's transition diagram. The second launches the \fsm{} visualization tool. The optional two-lists, \texttt{(s p)}, contain a state of the given machine and an invariant predicate for the state. Machine execution may always be visualized if the machine is a \dfa{}. Similarly to \texttt{sm-showtransitions}, \ndfa{} machine execution may only be visualized if the given word is in the machine's language. For further details on machine execution visualization in \fsm{}, the reader is referred to a previous publication \cite{Mor8}.

\begin{figure}[t!]
\begin{alltt}
     #lang fsm

     ;; L= ab* U ba*
     (define ab*Uba*-ndfa (make-ndfa \quot{}(S A B D E)
                                     \quot{}(a b)
                                     \quot{}S
                                     \quot{}(B C E F)
                                     \qquot{}((S ,EMP A) (S ,EMP D)
                                       (A a B) (B b B)
                                       (D b E) (E a E))))

     (check-equal? (sm-apply ab*Uba*-ndfa \quot{}(b b))   \quot{}reject)
     (check-equal? (sm-apply ab*Uba*-ndfa \quot{}(a a b)) \quot{}reject)
     (check-equal? (sm-apply ab*Uba*-ndfa \quot{}(a))     \quot{}accept)
     (check-equal? (sm-apply ab*Uba*-ndfa \quot{}(b))     \quot{}accept)
     (check-equal? (sm-apply ab*Uba*-ndfa \quot{}(a b b)) \quot{}accept)
     (check-equal? (sm-apply ab*Uba*-ndfa \quot{}(b a))   \quot{}accept)
\end{alltt}
\caption{The \fsm{} \ndfa{} for \texttt{L=\{$\texttt{ab}^{\texttt{*}} \cup \texttt{ba}^{\texttt{*}}$\}}.} \label{ndfa-fsm-ex}
\end{figure}

To illustrate programming finite-state machines in \fsm{}, consider the \ndfa{} to decide \texttt{L=\{$\texttt{ab}^{\texttt{*}} \cup \texttt{ba}^{\texttt{*}}$\}} displayed in \Cref{ndfa-fsm-ex}. At the beginning, the machine nondeterministically decides if the given word is in \texttt{$ab^{\texttt{*}}$} or in \texttt{$ba^{\texttt{*}}$} and transitions, respectively, to \texttt{A} or \texttt{D}. The unit tests illustrate words that are and that are not in \texttt{L}. Observe that the programmer only specifies nondeterministic behavior (the transitions out of \texttt{S}) and is not burdened with implementing nondeterministic behavior.

\section{Overall Visualization Design}
\label{viz-design}

To reduce the extraneous cognitive load, \fsm{}'s visualizations generate and collect images for each transformation step. As part of each image, there is a brief informative message that explains the step taken. The user only needs to use the arrow keys to step through the transformation. The use of these keys is specified as follows:
\begin{alltt}
  \(\rightarrow\) Move to next visualization step   \(\leftarrow\) Move to previous visualization step
  \(\downarrow\)  Move to visualization's end       \(\uparrow\)  Move to visualization's start
\end{alltt}
The visualization always displays instructions for the use of the arrow keys.

\begin{figure}[t!]
\captionsetup[subfigure]{justification=centering}
\centering
\begin{subfigure}[b]{0.70\textwidth}
\centering
\includegraphics[scale=0.27]{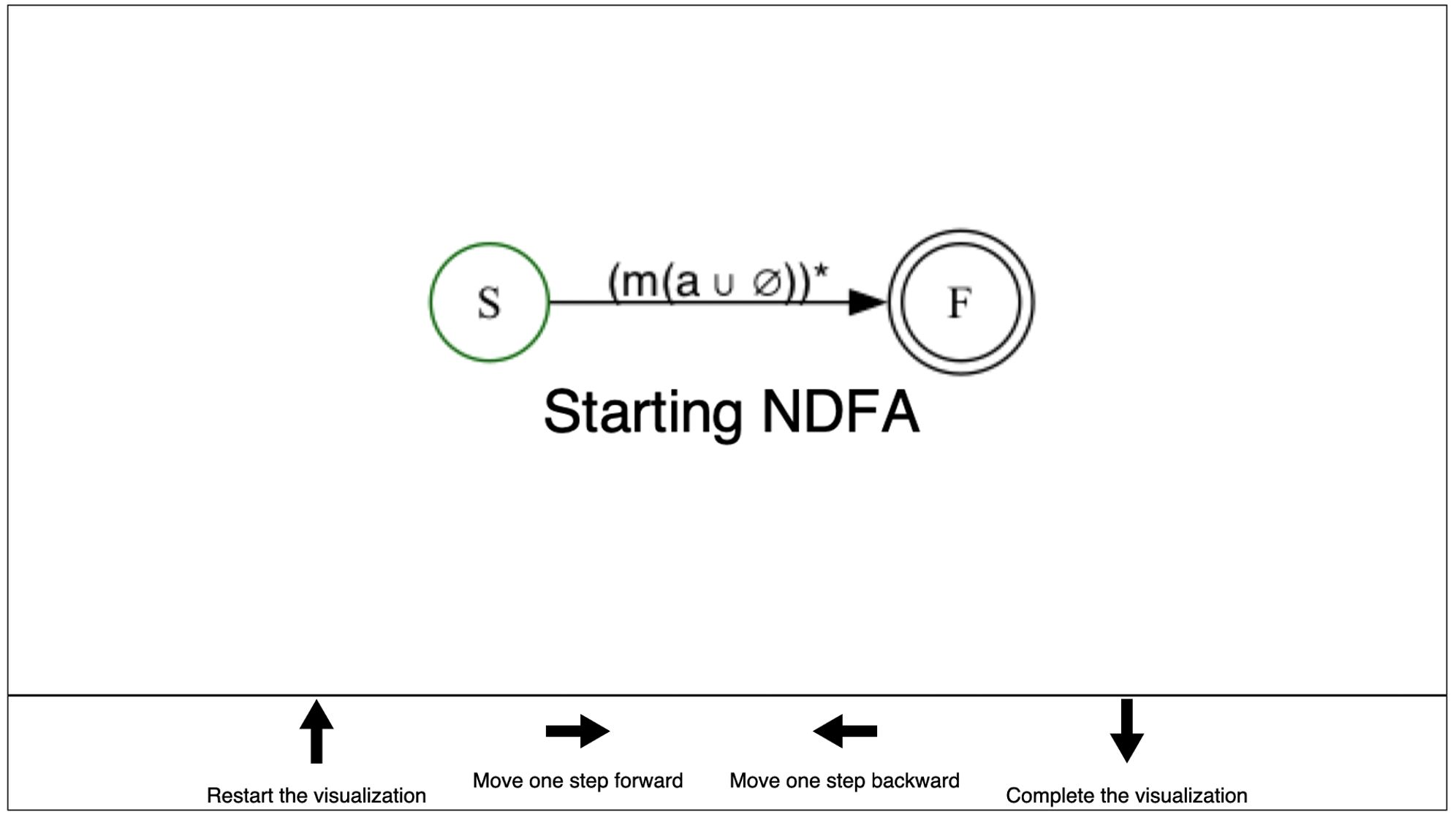}
\caption{Initial state for \texttt{(m(a $\cup$ $\varnothing$))$^*$}.} \label{initial-grph}
\end{subfigure}
\hfill
\begin{subfigure}[b]{0.70\textwidth}
\centering
\includegraphics[scale=0.28]{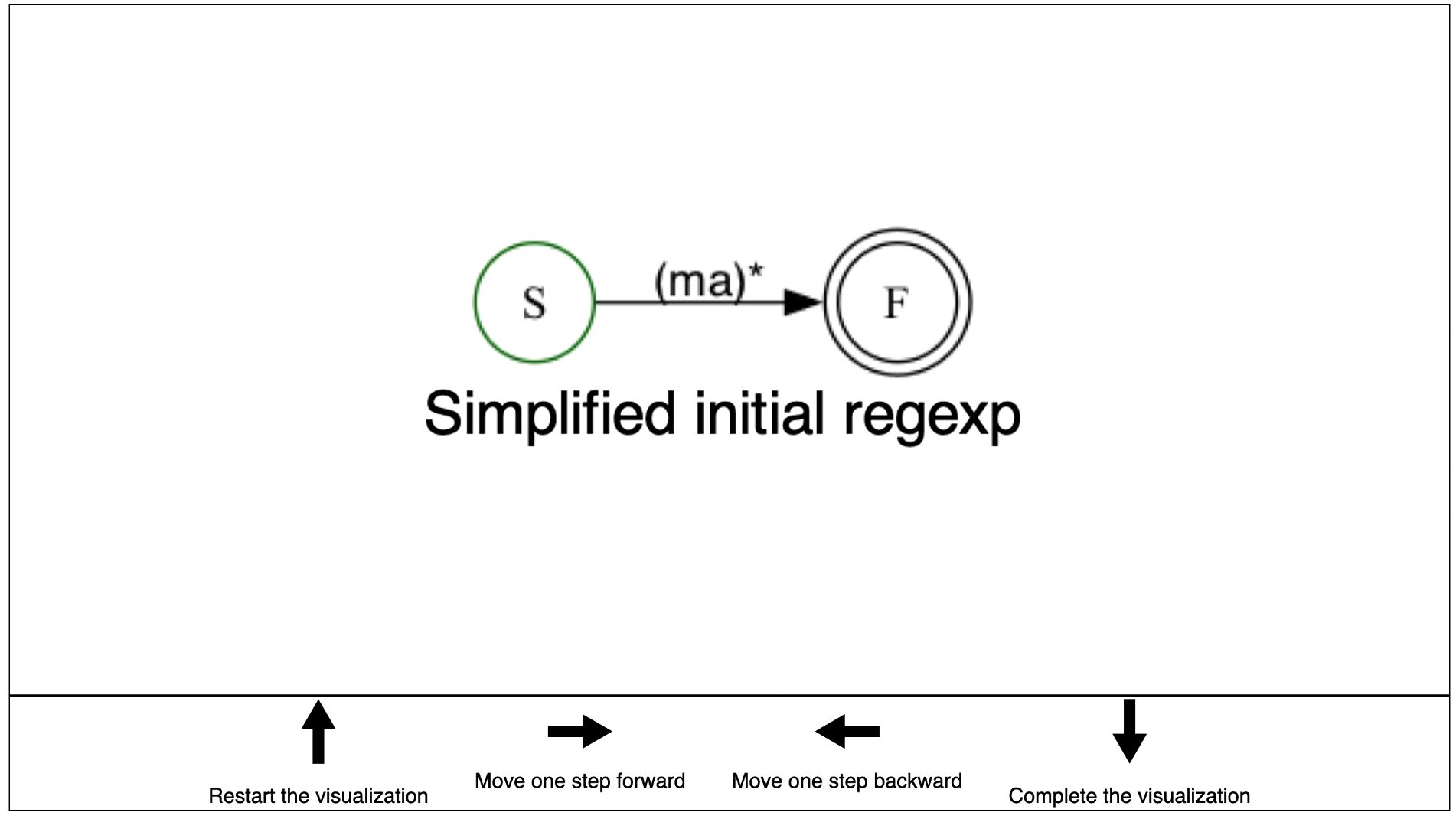}
\caption{State after simplifying the \regexp{}.}
\label{step1-grph}
\end{subfigure}
\caption{First visualization states in the \regexp{} to \ndfa{} transformation.}
\end{figure}

The images are stored in a structure, \texttt{viz-state}, that is defined as follows:
\begin{alltt}
     ;; A structure, (viz-state (listof images) (listof images)),
     ;; containing the processed and unprocessed images.
     (struct viz-state (pimgs upimgs))
\end{alltt}
The first list, \texttt{pimgs}, denotes the images previously displayed. The second list, \texttt{upimgs}, denotes the images to be displayed. The first image in \texttt{upimgs} is the currently displayed image. Initially, all images are in \texttt{upimgs}. Using the right arrow moves the first image from \texttt{upimgs} to \texttt{pimgs} and using the left arrow does the opposite. Using the down arrow moves all images from \texttt{upimgs}, except the last one, to \texttt{pimgs}. Using the up arrow moves all images to \texttt{upimgs}. Every time a step forward or backwards is taken, an informative message is placed at the bottom of each graphic along with arrow-use instructions. When appropriate, color is used to highlight the changes in the transformation. 

\subsection{Illustrative Example: \regexp{} to \ndfa{}}

\begin{figure}[t!]
\captionsetup[subfigure]{justification=centering}
\centering
\begin{subfigure}[b]{0.70\textwidth}
\centering
\includegraphics[scale=0.25]{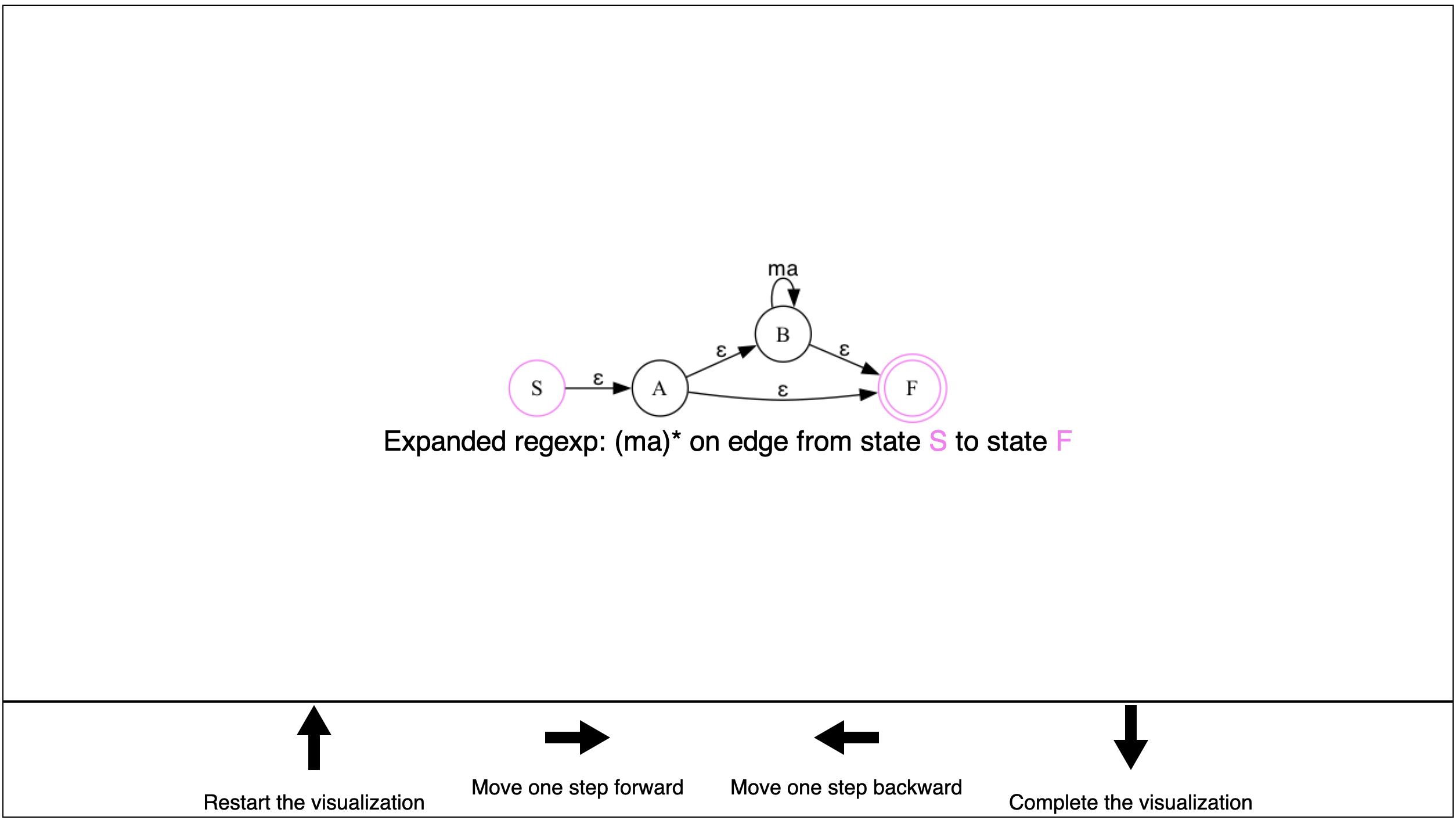}
\caption{Visualization state after expanding \texttt{(ma)$^{\texttt{*}}$}.} \label{kleene-expansion}
\end{subfigure}
\hfill
\begin{subfigure}[b]{0.70\textwidth}
\centering
\includegraphics[scale=0.25]{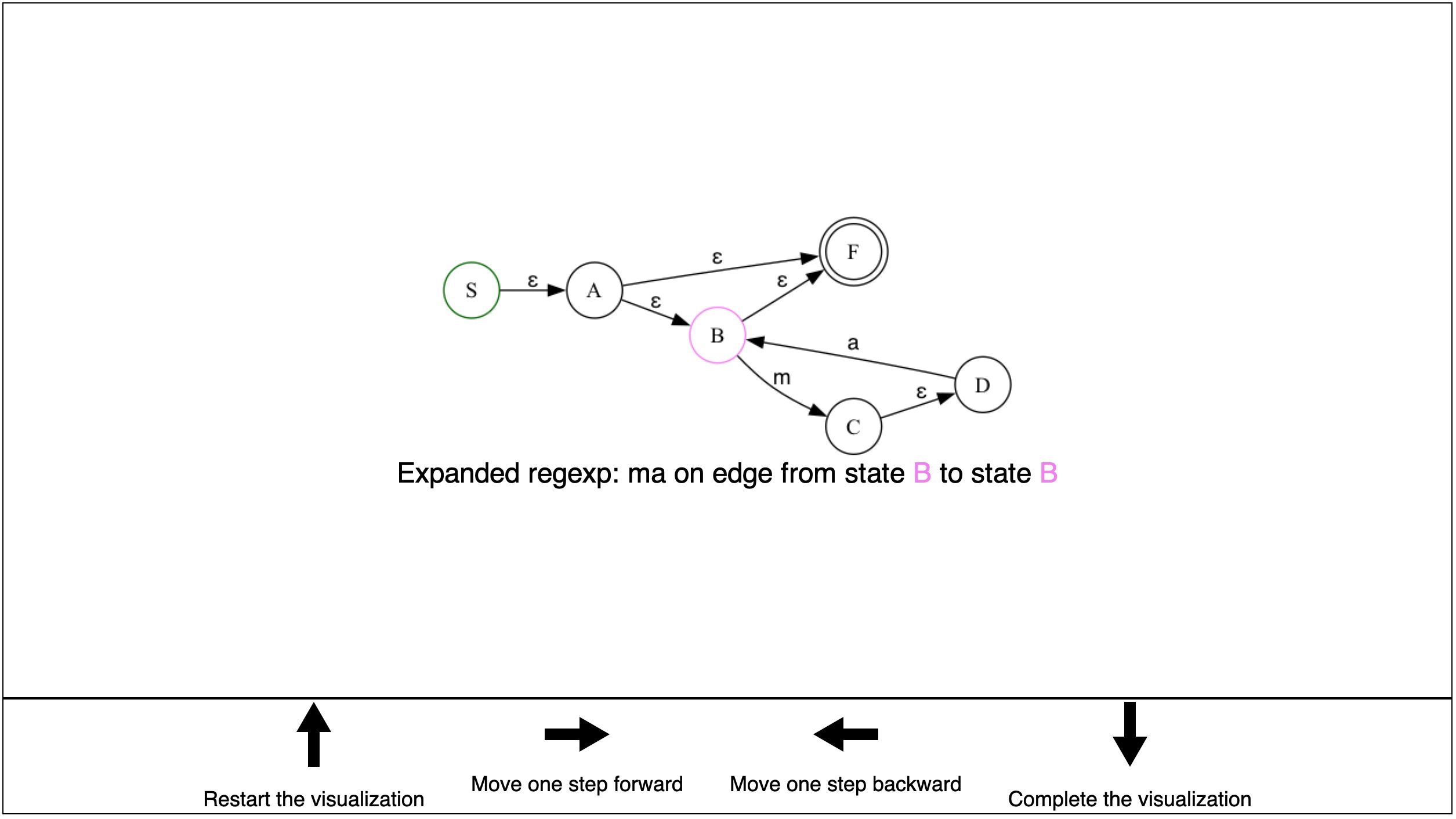}
\caption{Visualization state after expanding \texttt{ma}.}
\label{ma-expansion}
\end{subfigure}
\caption{Final visualization states in the \regexp{} to \ndfa{} transformation.}
\end{figure}

For illustrative purposes, consider the transforming \texttt{(m(a $\cup$ $\varnothing$))$^*$} into an \ndfa{}. The visualization's initial \gnfa{} is displayed in \Cref{initial-grph}. It contains a single transition from the starting state to the final state labeled with the \regexp{} to transform. The message indicates that it is the starting (approximation of the) \ndfa{}. The first step simplifies the given regular expression to \texttt{(ma)$^*$}. The visualization's state after this step is displayed in \Cref{step1-grph}. The message informs the user that the initial regular expression has been simplified. Technically, this step is not necessary but is useful to make the visualization more comprehensible for students that tend to write overly complex \regexp{}s. Next, the Kleene star \regexp{} that takes the machine from state \texttt{S} to state \texttt{F} is transformed. The visualization's state after this step is displayed in \Cref{kleene-expansion}. Observe that the message indicates the regular expression expanded, and the source and destination states. In both the message and in the graphic these states are highlighted in violet. The final step in the transformation expands \texttt{ma}. Given that this regular expression is on \texttt{B}'s self-transition, the source and destination states are the same. The state of the visualization after this expansion is displayed in \Cref{ma-expansion}. Observe that the message indicates the \regexp{} expanded and highlights in violet a single state.

At any point in the transformation, the user may move backwards in the transformation to examine before and after visualization states. This feature, along with the provided messages, allows the user to examine closely how the transformation is advanced by each step.

\subsection{Illustrative Example: \ndfa{} to \regexp{}}

\begin{figure}[t!]
\captionsetup[subfigure]{justification=centering}
\centering
\begin{subfigure}[b]{0.70\textwidth}
\centering
\includegraphics[scale=0.25]{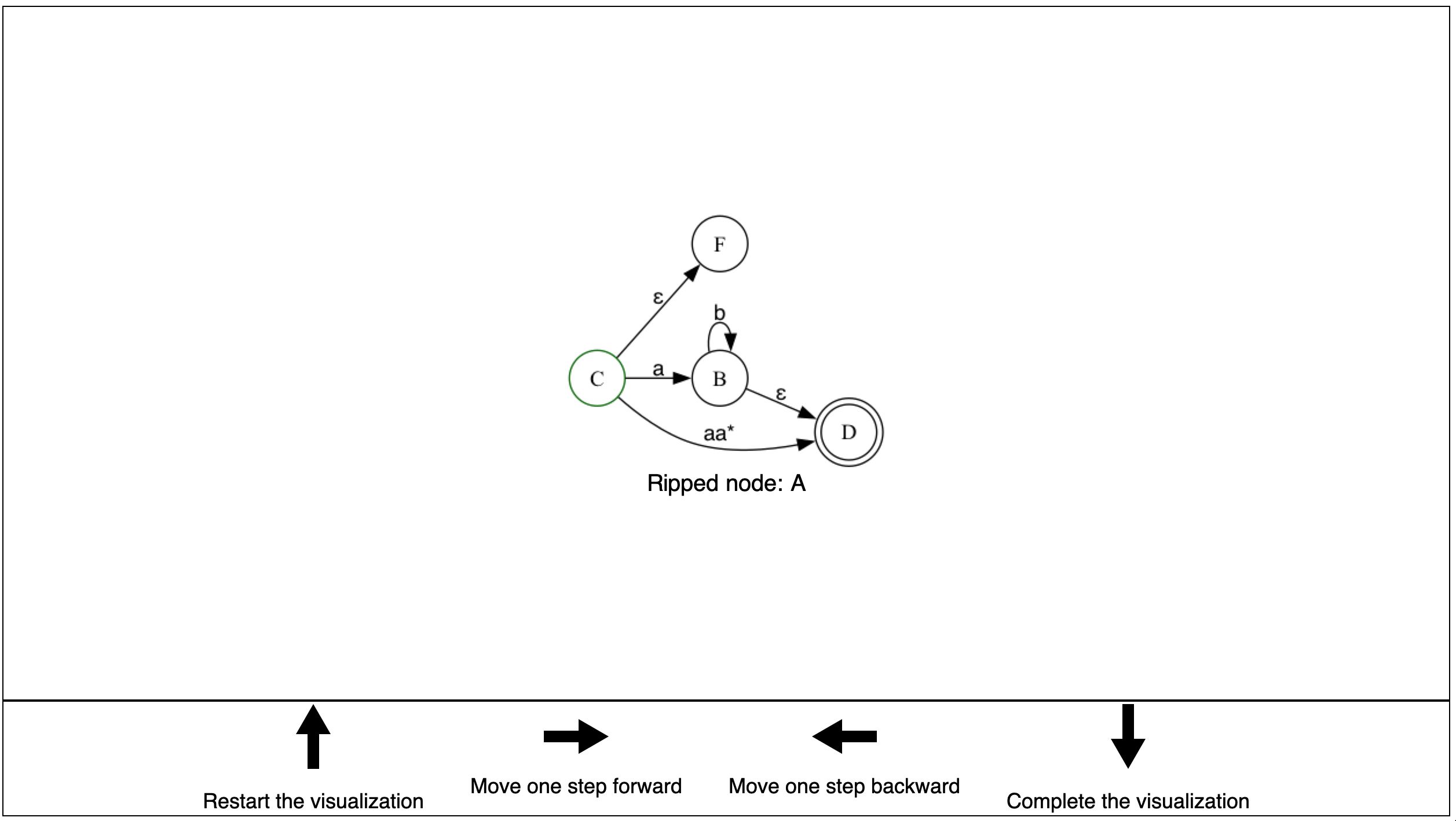}
\caption{Visualization state after ripping out \texttt{S} and \texttt{A}.}
\label{ripped-A}
\end{subfigure}
\hfill
\begin{subfigure}[b]{0.70\textwidth}
\includegraphics[scale=0.25]{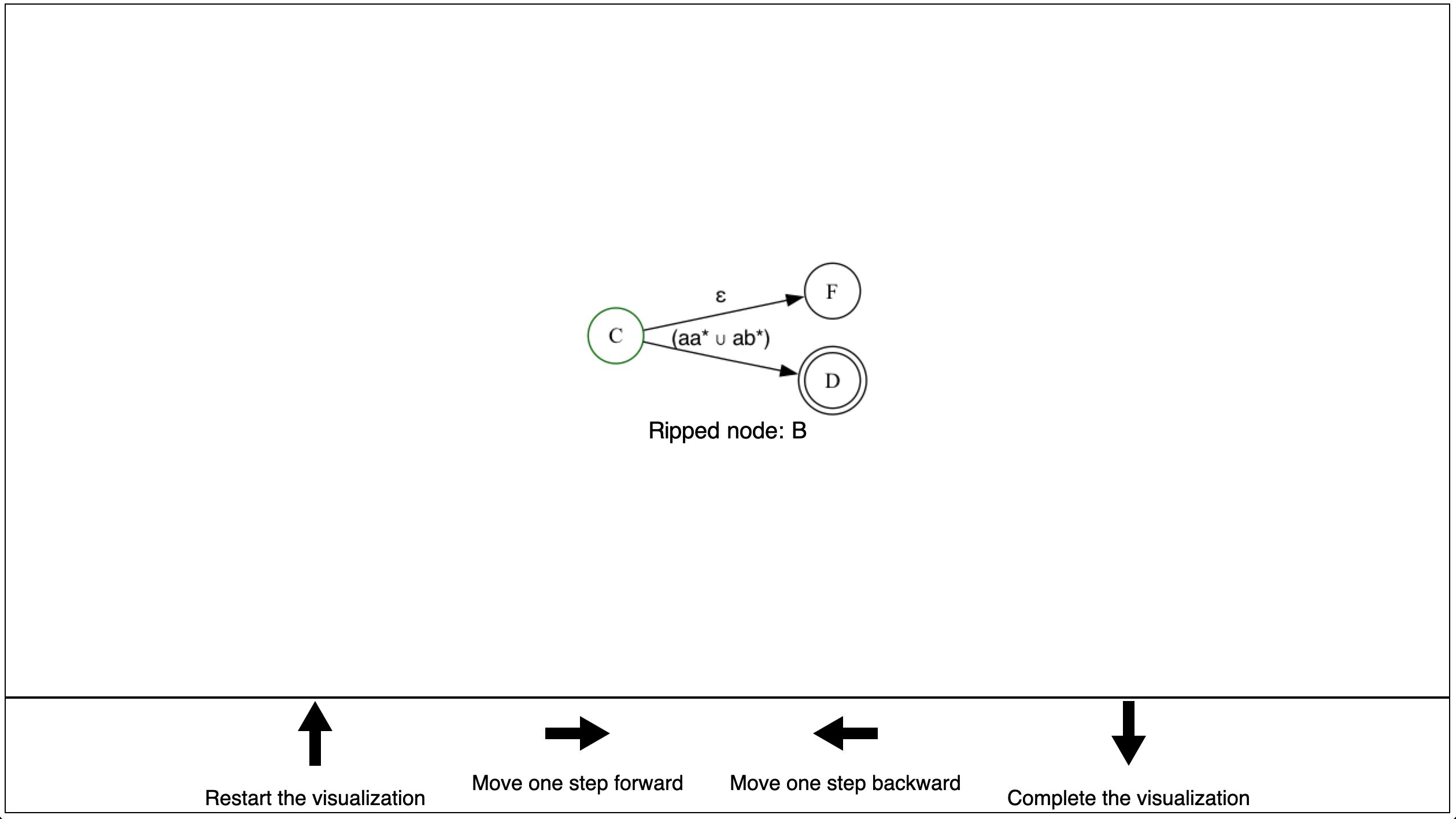}
\caption{Visualization state after ripping out \texttt{B}.} \label{ripped-B}
\end{subfigure}
\caption{Node-ripping visualization steps in the \ndfa{} to \regexp{} transformation.}
\end{figure}

For illustrative purposes, consider transforming the following \ndfa{}:
\begin{center}
\includegraphics[scale=0.35]{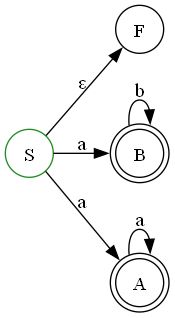}
\end{center}
The programmer has, unnecessarily, included a nonfinal state, \texttt{F}, that is only reachable by an empty transition from the starting state and that does not have any outgoing transitions. The visualization steps rip out a node one at a time. \Cref{ripped-A} displays the visualization state after ripping out \texttt{S} and \texttt{A}. Observe that there is a transition on \texttt{aa$^{\texttt{*}}$} from \texttt{C} to \texttt{D} resulting from ripping out the two nodes. Ripping out \texttt{B} means a new transition is needed from \texttt{C}, the only predecessor, to, \texttt{D}, the only successor. This results in two edges between \texttt{C} and \texttt{D} and, thus, they are consolidated using a union regular expression resulting in the visualization state displayed in \Cref{ripped-B}. Finally, ripping out \texttt{F} has no effect on the edge from \texttt{C} and \texttt{D}, which is labeled with the resulting regular expression.

At any point in the transformation, the user may move backwards to examine before and after visualization states. Thus, allowing the student to closely examine how nodes are ripped out and new transitions are created. 

\section{Implementation}
\label{impl}

\subsection{Constructing the Graphics for the \regexp{} to \ndfa{} Transformation}
\label{regexp2ndfa}

During the transformation, there exists a \gnfa{} whose transitions are labeled with arbitrary \regexp{}s. The goal is to transform the \gnfa{} so that its transitions are only labeled with singleton and empty \regexp{}s. At each step, a transition labeled with a union, a concatenation, or a Kleene star \regexp{} is chosen to be transformed. These transformations are based on well-known constructors for closure properties of regular languages \cite{Lewis,PBFLAT,Rich,Sipser}. When a \regexp{} is transformed, the chosen transition is removed from the \gnfa{} and new states and edges are added. \gviz{} is used to generate a new graphic.

A union \regexp{}, \texttt{(union-regexp r$_{\texttt{1}}$ r$_{\texttt{2}}$)}, labeling the transition between two states \texttt{S} and \texttt{F} is transformed by creating four fresh states: say, \texttt{A}, \texttt{B}, \texttt{C}, and \texttt{D}. Each branch of the union exclusively uses two of these states and they are connected by a transition labeled with the corresponding regular expression for the branch. For instance, \texttt{A}, and \texttt{C} are connected using \texttt{r$_{\texttt{1}}$} and \texttt{B}, \texttt{D} are connected using \texttt{r$_{\texttt{2}}$}. \texttt{S} is connected to \texttt{A} and \texttt{B} by empty transitions. \texttt{C} and \texttt{D} are connected to \texttt{F} by empty transitions. Visually, transforming:
\begin{center}
\includegraphics[scale=0.5]{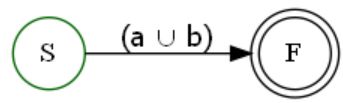}
\end{center}
results in:
\begin{center}
\includegraphics[scale=0.38]{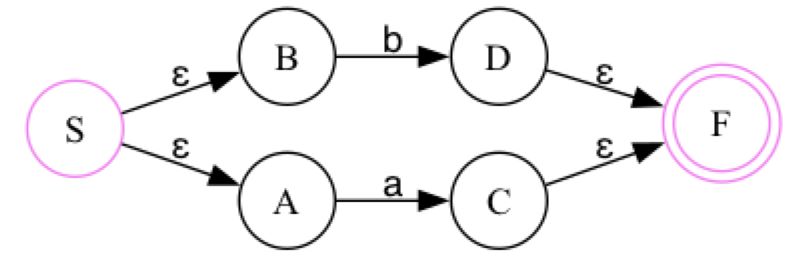}
\end{center}
Finally, \texttt{S} and \texttt{F} are highlighted in violet indicating the head and the tail of the replaced edge. 

A concatenation \regexp{}, \texttt{(concat-regexp r$_{\texttt{1}}$ r$_{\texttt{2}}$)}, labeling the transition between two states \texttt{S} and \texttt{F} is transformed by creating two fresh states: say, \texttt{A} and \texttt{B}. A transition from \texttt{S} to \texttt{A} labeled with \texttt{r$_{\texttt{1}}$}, a transition from \texttt{A} to \texttt{B} labeled with an empty \regexp{}, and a transition from \texttt{B} to \texttt{F} labeled with r$_{\texttt{2}}$ are added to the \gnfa{}. Visually, transforming:
\begin{center}
\includegraphics[scale=0.5]{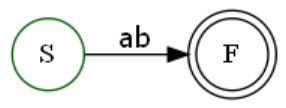}
\end{center}
results in:
\begin{center}
\includegraphics[scale=0.38]{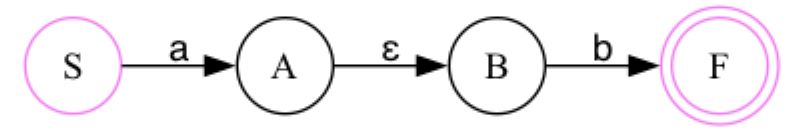}
\end{center}
Finally, \texttt{S} and \texttt{F} are highlighted in violet indicating the head and the tail of the replaced edge. 

A Kleene star \regexp{}, \texttt{(kleenestar-regexp r$_{\texttt{1}}$)}, labeling the transition between two states \texttt{S} and \texttt{F} is transformed by creating two fresh states: say, \texttt{A} and \texttt{B}. \texttt{S} is connected to \texttt{A}, \texttt{A} is connected to \texttt{B}, \texttt{A} is connected to \texttt{F}, and \texttt{B} is connected to \texttt{F} by transitions labeled with an empty \regexp{}. Finally, there is a loop transition on \texttt{B} labeled with \texttt{r$_{\texttt{1}}$}. Visually, transforming:
\begin{center}
\includegraphics[scale=0.5]{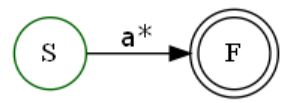}
\end{center}
results in:
\begin{center}
\includegraphics[scale=0.38]{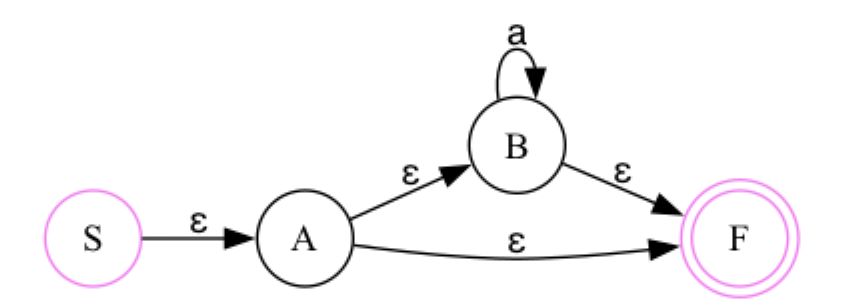}
\end{center}
Finally, \texttt{S} and \texttt{F} are highlighted in violet indicating the head and the tail of the replaced edge. 

\subsection{Constructing the Graphics for the \ndfa{} to \regexp{} Transformation}
\label{ndfa2regexp}

The bulk of the graphics are created by ripping out nodes. Ripping out a node \texttt{A} requires the removal of transitions into and out of \texttt{A} and the generation of new transitions connecting each predecessor of \texttt{A} with each successor of \texttt{A}. There are two cases that need to be distinguished: either \texttt{A} has or does not have a loop transition on it.

If \texttt{A} does not have a loop on it then each predecessor of \texttt{A} is connected to each successor of \texttt{A} by a transition labeled with a concatenation \regexp{} that contains the regular expression from the predecessor to \texttt{A} and the regular expression from \texttt{A} to the successor. For instance, if \texttt{(M r$_{\texttt{1}}$ A)} and \texttt{(A r$_{\texttt{2}}$ N)} are transitions in the current \gnfa{} then these two transitions are removed and substituted with \texttt{(M r$_{\texttt{1}}$r$_{\texttt{2}}$ N)}. Visually, if the current \gnfa{} is:
\begin{center}
\includegraphics[scale=0.5]{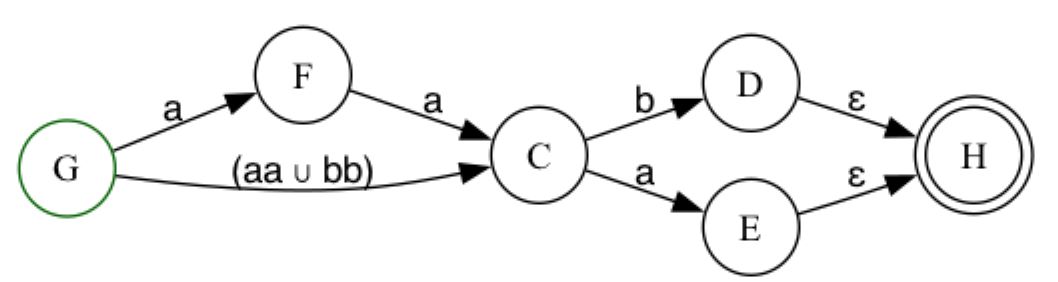}
\end{center}
Ripping out \texttt{C} means that \texttt{G} and \texttt{F} must be connected to \texttt{E} and \texttt{D}. The new edges generated are labeled with the concatenation of each edge into \texttt{C} and each edge out of \texttt{C}. The resulting \gnfa{} is:
\begin{center}
\includegraphics[scale=0.5]{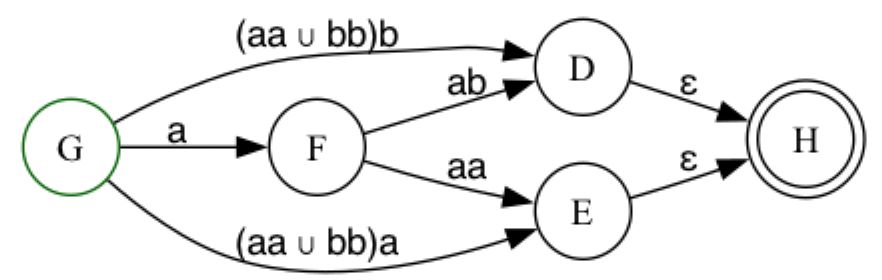}
\end{center}

If \texttt{A} has a loop on itself then each predecessor of \texttt{A} is connected to each successor of \texttt{A} by a transition labeled with the concatenation of the regular expression from the predecessor to \texttt{A}, a Kleene star regular expression for the loop's regular expression, and the regular expression to the successor of \texttt{A}. For instance, if \texttt{(M r$_{\texttt{1}}$ A)}, \texttt{(A r$_{\texttt{2}}$ A)} and \texttt{(A r$_{\texttt{3}}$ N)} are transitions in the current \gnfa{} then, when \texttt{A} is ripped out, these three transitions are removed and substituted with \texttt{(M r$_{\texttt{1}}$r$_{\texttt{2}}^{\texttt{*}}$r$_{\texttt{3}}$ N)}. Visually, if the current \gnfa{} is:
\begin{center}
\includegraphics[scale=0.5]{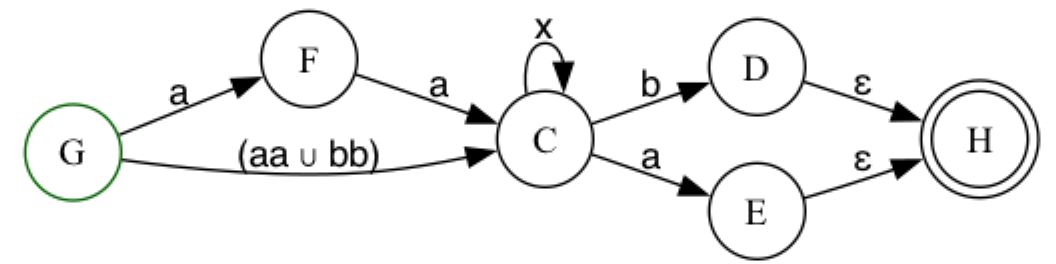}
\end{center}
Ripping out \texttt{C} means that \texttt{F} and \texttt{G} must have transitions to \texttt{E} and \texttt{D}. The new edges generated are labeled with the concatenation of each edge into \texttt{C}, a Kleene star regular expression containing the label on \texttt{C}'s loop, and the edges out of \texttt{C}. The resulting \gnfa{} is:
\begin{center}
\includegraphics[scale=0.5]{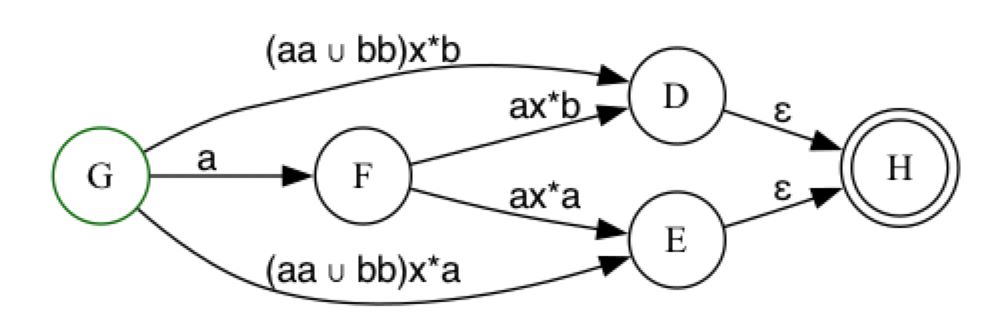}
\end{center}

\section{Empirical Data}
\label{study}

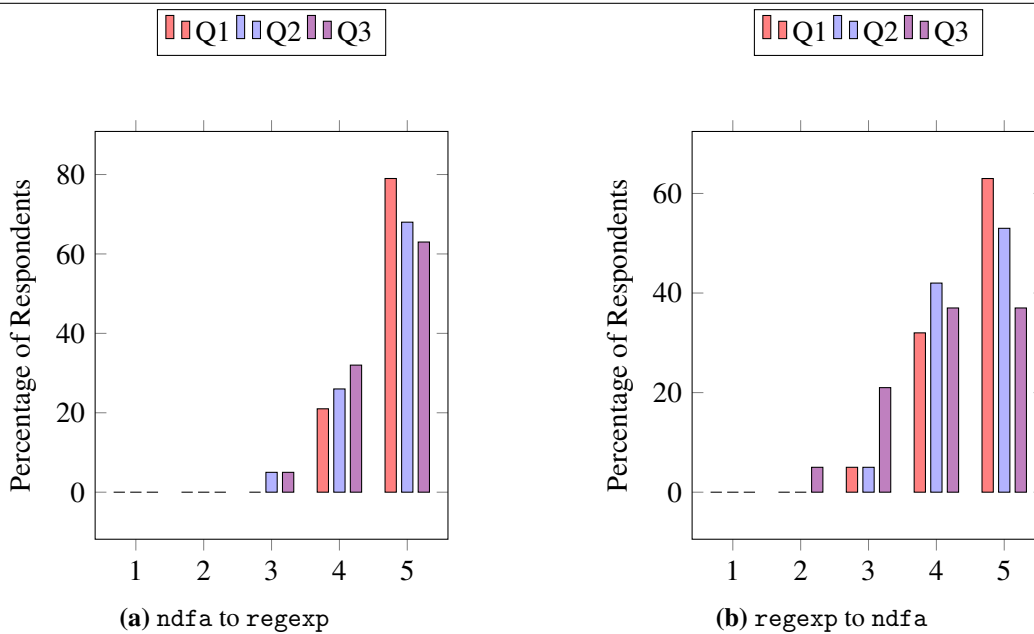
\begin{figure}[t!]
\captionsetup[subfigure]{justification=centering}
\centering
\begin{subfigure}[b]{0.49\textwidth}
\centering
\begin{tikzpicture}[scale=1.0]
\begin{axis}[
    bar width=.15cm,
    width=0.8\textwidth,
    height=.32\textheight,
    ybar,
    enlargelimits=0.15,
    legend style={at={(0.5,1.3)},
      anchor=north,legend columns=-1},
    ylabel={Percentage of Respondents},
    symbolic x coords={1,2,3,4,5},
    xtick=data,
    nodes near coords align={vertical},
    ]
\addplot[color=black,fill=red!50] coordinates {(1,0) (2,0) (3,0) (4,21) (5,79)};
\addplot[color=black,fill=blue!30] coordinates {(1,0) (2,0) (3,5) (4,26) (5,68)};
\addplot[color=black,fill=violet!50] coordinates {(1,0) (2,0) (3,5) (4,32) (5,63)};
\legend{Q1, Q2, Q3}
\end{axis}
\end{tikzpicture}
\caption{\ndfa{} to \regexp{}} \label{emp-data1}
\end{subfigure}
\begin{subfigure}[b]{0.49\textwidth}
\centering
\begin{tikzpicture}[scale=1.0]
\begin{axis}[
    bar width=.15cm,
    width=.8\textwidth,
    height=.32\textheight,
    ybar,
    enlargelimits=0.15,
    legend style={at={(0.5,1.3)},
      anchor=north,legend columns=-1},
    ylabel={Percentage of Respondents},
    symbolic x coords={1,2,3,4,5},
    xtick=data,
    nodes near coords align={vertical},
    ]
\addplot[color=black,fill=red!50] coordinates {(1,0) (2,0) (3,5) (4,32) (5,63)};
\addplot[color=black,fill=blue!30] coordinates {(1,0) (2,0) (3,5) (4,42) (5,53)};
\addplot[color=black,fill=violet!50] coordinates {(1,0) (2,5) (3,21) (4,37) (5,37)};
\legend{Q1, Q2, Q3}
\end{axis}
\end{tikzpicture}
\caption{\regexp{} to \ndfa{}} \label{emp-data2}
\end{subfigure}
\caption{Control group response distribution.} \label{emp-data}
\end{figure}

To initially assess the usefulness of our new teaching tools, before deploying them in a classroom setting, we collected empirical data from a focus group using a voluntary anonymous survey\footnote{None of the volunteers received any benefits for their participation.}. Nineteen students volunteered to participate (5 Seton Hall undergraduates who have not yet taken a  \texttt{FLAT} class; 13 Instituto Universit\'{a}rio de Lisboa undergraduates currently taking a \texttt{FLAT} class, and 1 graduate student that works as a \texttt{FLAT} teaching assistant at Instituto Universit\'{a}rio de Lisboa)\footnote{Only the 5 volunteers from Seton Hall University are familiar with \racket{}-like languages (specifically, the \racket{} student languages used in \cite{HtDP2,APS,APD})}. They were all introduced to the \ndfa{} to \regexp{} and to the \regexp{} to \ndfa{} transformations using the \fsm{} tools described in this article. Prior to learning about the transformations, students got a brief introduction to \fsm{} focusing on programming \ndfa{}s and \regexp{}s. After learning about each of the transformations and using the visualization tools, the students took a survey with the following questions about the \ndfa{} to \regexp{} visualization tool:
\begin{alltt}
     \textrm{\textcolor{red!50}{Q1}: Overall, how useful is the visualization to understand the ndfa to regexp transformation?}
     \textrm{\textcolor{blue!30}{Q2}: How difficult is it to use the visualization?}
     \textrm{\textcolor{violet!50}{Q3}: How difficult is it to understand a visualized transformation?}
\end{alltt}
Respondents answered using a Likert scale \cite{Likert}. Question 1 uses the scale from [1] Not at all useful to [5] Extremely useful. Questions 2 and 3 use the scale from [1] Extremely difficult to [5] Extremely easy. The distribution of responses is displayed in \Cref{emp-data1}. 

Responses to \texttt{Q1} indicate that all respondents feel that the visualization tool is useful to understand the transformation from \ndfa{} to \regexp{} (responses 4 and 5). These results were not anticipated given that most respondents have little to no experience with formal languages and automata theory. This suggests that the visualization is useful even for \flatt{} beginners.

Responses to \texttt{Q2} indicate that most respondents, 94\%, feel strongly that the visualization tool is easy to use (responses 4 and 5). This suggests that the efforts made to reduce the extraneous cognitive load associated with learning how to use the visualization are successful.

Responses for \texttt{Q3} indicate that most respondents, 95\%, feel strongly that the visualized transformation is easy to understand. This is also an unexpected result given that most respondents were not familiar with the transformation algorithms. It suggests that the size of each step in the visualization makes the transformation accessible to novices.

The second part of the survey addressed the \regexp{} to \ndfa{} transformation. The survey included the following questions:
\begin{alltt}
     \textrm{\textcolor{red!50}{Q1}: Overall, how useful is the visualization to understand the regexp to ndfa transformation?}
     \textrm{\textcolor{blue!30}{Q2}: How difficult is it to use the visualization?}
     \textrm{\textcolor{violet!50}{Q3}: How difficult is it to understand a visualized transformation?}
\end{alltt}
These questions are also answered using a Likert scale \cite{Likert}. Question 1 uses the scale from [1] Not at all useful to [5] Extremely useful. Questions 2 and 3 use the scale from [1] Extremely difficult to [5] Extremely easy. The distribution of responses is displayed in \Cref{emp-data2}.

For \texttt{Q1}, we observe that respondents feel strongly, with 95\% answering 4 or 5, that the visualization tool is useful to understand the transformation from \regexp{} to \ndfa{}. These results are unexpected as most respondents, as observed earlier, have no prior experience with formal languages and automata theory. Along with the results obtained for \texttt{Q1} for the previous transformation above, this suggests that providing a visual trace of construction algorithms benefits students at all levels of experience.

For \texttt{Q2}, we observe that most respondents, 95\%, feel strongly that the visualization is easy to use. This suggest that our efforts to keep the extraneous cognitive load low are successful. We attribute this to the easy-to-use arrow-key interface and the informative messages at each step.

For \texttt{Q3}, we observe that a majority of respondents, 74\%, feel that the transformation is easy to understand (answers 4 and 5). A significant minority of respondents, 21\%, felt less strongly (response 3). Such a distribution is expected among students beginning in \flatt{} given that, to fully understand this transformation, the respondents need to be familiar with closure properties for regular languages and the corresponding construction algorithms. Nonetheless, these results are very encouraging given that even the novices felt they understood the transformation.  

In addition, the respondents were asked qualitative questions. The following responses were obtained when respondents were asked about their favorite characteristics of the visualization tools:
\begin{alltt}
     "Being able to cycle step-by-step through each step in the 
      visualization is super useful. I'm thinking of how my students 
      might not understand a particular step, and I can just cycle 
      back and forth as much as I need :) I also really liked the 
      messages that explain what was done in each step, feel like 
      they really help to keep track of what's going on!"

     "It makes it much easier to juggle all the different states in 
      my head."

     "I like the purple highlight coloring of the node that is to be 
      broken down."

     "They are pretty straightforward and easy to understand.
      The colors are nice and easy to follow too."

     "Really easy to understand what's going on. I will, for sure,
      use it for studying."

\end{alltt}
This feedback suggests that, due to the perceived clarity and the readability, students in a course setting will welcome and use the visualization tools.

The following responses were given when asked what they liked the least about the visualization tools:
\begin{alltt}
     "The long I-xyzw... state names can make the visualization
     somewhat overwhelming in my opinion."

     "Maybe the name of the new states should have a better 
      naming scheme instead of random names"

     "The movement of nodes instead of static points and growing frame."

\end{alltt}
The first two comments refer to the prior names randomly generated for new states. The prior names included a random 6-digit natural number (e.g., \texttt{I-872431}). In light of the above feedback and prior to publication, random state-name generation has been updated to only include, if necessary, a random number. The new generation technique produces the shortest possible state-name not in use in the construction of an \ndfa{}. The use of this new random state-name generation is reflected in the previous sections of this article (i.e., a random state-name with a 6-digit natural number is not generated for any of the examples used).

The second concern refers to the placement of nodes in the generated graphs. Given that drawing graphs is complex, the main impetus for current research on computer-aided graph drawing is to facilitate the visual analysis of various kinds of complex networked or connected systems \cite{Kruja}. Several graph drawing libraries have been built and successfully deployed. Among the most widely used is \gviz{} \cite{graphviz} and \fsm{} uses \gviz{} to generate its diagrams. \gviz{}, however, provides no control over node placement and we must accept state movement as diagrams grow.

\section{Concluding Remarks}
\label{concls}

This article presents novel \fsm{} visualization tools for an \ndfa{} to \regexp{} transformation and for a \regexp{} to \ndfa{} transformation. The visualizations simultaneously render the transition diagram images and display informative messages to assist the user navigate the transformation. It improves the previous approaches by rendering transition diagrams in an appealing manner. In addition, the \regexp{} to \ndfa{} visualization tool has appropriate state color coding to clearly illustrate which edge has been expanded. The \fsm{} visualization tools, unlike any other visualization tools for these transformations, can advance both forwards and backwards. Finally, both transformations can be completed in a single step, or restarted in a single click from any point in the computation, without preventing the user from moving the simulation forwards and backwards. All these advancements are done by clicking the arrow keys. Thus, helping lower the extraneous cognitive load associated with learning how to use the tools.

Future work includes using the described tools in a classroom setting and measuring student impressions. We envision using the visualizations to help students understand formal statements. That is, the plan is to introduce students to the transformation algorithms using formal notation (given that it is important for students to understand formal statements) and, in tandem, to use the visualization tools to help students understand the formal notation so that they can implement the algorithms in \fsm{}. Future work also includes developing visualization tools for construction algorithms based on closure properties for regular languages including union, concatenation, Kleene star, complement, and intersection. In addition, we are expanding the reach of our visualization tools into derivations for regular, context-free, and context-sensitive grammars. The goal is to assist students understand why a word is a member of a language through the creation of parse trees.

\subsubsection*{Acknowledgements.}
The authors thank Filipe Alexandre Azinhais dos Santos and Alfonso Manuel Barral Caniço from Instituto Universit\'{a}rio de Lisboa for inviting us to their classroom to conduct our control group study. In addition, the authors thank Oliwia Kempinski, Andr\'{e}s Maldonado, Josie Des Rosiers, and Shamil Dzhatdoyev for their feedback on previous versions of this manuscript.

\bibliographystyle{eptcs}
\bibliography{fsa-regexp-viz}
\end{document}